\input{aipcheck}

\documentclass[
    ,final            
  ]
  {aipproc}

\layoutstyle{6x9}

\begin{document}

\title{Realistic shell-model calculations and exotic nuclei around $^{132}$Sn}

\classification{21.60.Cs, 21.30.Fe, 27.60.+j, 27.80.+w}

\keywords{Realistic shell model; $^{132}$Sn region; $^{208}$Pb region}

\author{A. Covello}{
  address={Dipartimento di Scienze Fisiche, Universit\`a di Napoli Federico II, and Istituto Nazionale di Fisica Nucleare, 
Complesso Universitario di Monte S. Angelo, I-80126 Napoli, Italy} }

\author{L. Coraggio } {
  address={Istituto Nazionale di Fisica Nucleare, Complesso Universitario di Monte S. Angelo, I-80126 Napoli, Italy} }

\author{A. Gargano} {
  address={Istituto Nazionale di Fisica Nucleare, Complesso Universitario di Monte S. Angelo, I-80126 Napoli, Italy} }

 \author{N. Itaco} {
 address={Dipartimento di Scienze Fisiche, Universit\`a di Napoli Federico II, and Istituto Nazionale di Fisica Nucleare, Complesso Universitario di Monte S. Angelo, I-80126 Napoli, Italy} }


\begin{abstract}
 We report on a study of exotic nuclei around doubly magic $^{132}$Sn in terms of the shell model employing a realistic effective interaction derived from the CD-Bonn nucleon-nucleon potential. The short-range repulsion of the latter is renormalized by constructing a smooth low-momentum potential, $V_{\rm low-k}$, that is used directly as input for the calculation of the effective interaction. In this paper, we focus attention on proton-neutron multiplets in the odd-odd nuclei $^{134}$Sb, $^{136}$Sb. We show that the behavior of these multiplets is quite similar to that of the analogous multiplets in  the counterpart nuclei in the $^{208}$Pb region, $^{210}$Bi and $^{212}$Bi.

\end{abstract}

\maketitle


\section{Introduction}

We have recently studied \cite{Coraggio05,Coraggio06,Covello07,Simpson07} several nuclei beyond doubly magic $^{132}$Sn within the framework of the shell model employing realistic effective interactions derived from the CD-Bonn nucleon-nucleon ($NN$) potential \cite{Machleidt01}. 
A main difficulty encountered in this kind of calculations is the strong short-range repulsion contained in the bare $NN$ potential $V_{NN}$, which prevents its direct use in the derivation of the shell-model effective interaction $V_{\rm eff}$.
As is well known, the traditional way  to overcome this difficulty is the Brueckner
$G$-matrix method. Instead, in the calculations mentioned above we have made use of a new approach  \cite{Bogner02} which consists in deriving from $V_{NN}$ a low-momentum potential, $V_{\rm low-k}$, that preserves the deuteron binding energy and scattering phase shifts of $V_{NN}$ up to a certain cutoff momentum $\Lambda$. This is a smooth potential which can be used directly to derive $V_{\rm eff}$, and it has been
shown \cite{Bogner02,Covello03} that it provides an advantageous alternative to the use of the $G$ matrix. 

In this paper, we shall focus attention on the proton-neutron multiplets in the two odd-odd Sb isotopes $^{134}$Sb and $^{136}$Sb, which are most appropriate for testing the matrix elements of the proton-neutron interaction between valence nucleons outside $^{132}$Sn. Note that $^{136}$Sb with an $N/Z$ ratio of 1.67 is at present the most exotic open-shell nucleus beyond $^{132}$Sn for which information exists on excited states. 

While it is very difficult to obtain information on neutron-rich nuclei around $^{132}$Sn, which lie well away from the valley of stability, this is not the case for the  $^{208}$Pb neighbors, which have long been the subject of experimental and theoretical studies. The new data which are becoming available for the $^{132}$Sn region make it possible to investigate more quantitatively the resemblance between the spectroscopy of this region and that of nuclei around $^{208}$Pb, which has been pointed out in several recent papers \cite{Fornal01,Isakov06,Korgul07}.
 
With the above motivation, we have calculated the proton-neutron multiplets in $^{210}$Bi and $^{212}$Bi, which are the counterparts of $^{134}$Sb and $^{136}$Sb in the Pb region. 

We start by giving a brief description of the theoretical framework in which our realistic shell-model calculations are performed and then present and discuss our results. A short summary is given in the last section.\\

\section{Outline of theoretical framework}

The shell-model effective interaction $V_{\rm eff}$ is defined, as usual, in the following way. In principle, one should solve a nuclear many-body Schr\"odinger equation of the form 
\begin{equation}
H\Psi_i=E_i\Psi_i ,
\end{equation}
with $H=T+V_{NN}$, where $T$ denotes the kinetic energy. This full-space many-body problem is reduced to a smaller model-space problem of the form
\vspace{-.1cm}
\begin{equation}
PH_{\rm eff}P \Psi_i= P(H_{0}+V_{\rm eff})P \Psi_i=E_iP \Psi_i .
\end{equation}
\noindent Here $H_0=T+U$ is the unperturbed Hamiltonian, $U$ being an auxiliary potential introduced to define a convenient single-particle basis, and $P$ denotes the projection operator onto the chosen model space.

As pointed out in the Introduction, we ``smooth out'' the strong repulsive core contained in the bare $NN$ potential $V_{NN}$ by constructing a low-momentum  potential $V_{\rm low-k}$. This is achieved by integrating out the high-momentum modes of $V_{NN}$ down to a cutoff momentum  $\Lambda$. This integration is carried out with the requirement that the deuteron binding energy and phase shifts of $V_{NN}$ up to $\Lambda$ are preserved by $V_{\rm low-k}$. 
A detailed description of the derivation of $V_{\rm low-k}$ from $V_{NN}$ as well as a discussion of its main features can be found in Refs. \cite{Bogner02,Covello05}. 

Once the $V_{\rm low-k}$ is obtained, we use it as input interaction  for the calculation of the matrix elements of the shell-model effective interaction. The latter is derived by employing a folded-diagram method, which was previously applied to many nuclei using $G$-matrix interactions \cite{Covello01}. Since $V_{\rm low-k}$ is already a smooth potential, it is no longer necessary to calculate the $G$ matrix. We therefore derive  $V_{\rm eff}$ following the same procedure as described, for instance, in Refs. \cite{Jiang92,Covello97}, except that the $G$ matrix used there is replaced by $V_{\rm low-k}$. More precisely, we first calculate the so-called $\hat{Q}$-box \cite{Kuo90} including diagrams up to second order in the two-body interaction. The shell-model effective interaction is then obtained by summing up the $\hat{Q}$-box folded diagram series using the Lee-Suzuki iteration method \cite{Suzuki80}.

\section{Calculations and results}

In our calculations for  $^{134}$Sb and $^{136}$Sb we assume that the valence
neutrons occupy the six levels $0h_{9/2}$, $1f_{7/2}$, $1f_{5/2}$, $2p_{3/2}$,
$2p_{1/2}$, and  $0i_{13/2}$ of the 82-126 shell, while for the odd proton the model space includes the five  levels  $0g_{7/2}$, $1d_{5/2}$, $1d_{3/2}$, $2s_{1/2}$, and $0h_{11/2}$ of the 50-82 shell. 
Similarly, for the two Bi isotopes we take as model space for the valence proton the six levels of the 82-126 shell and let the valence neutrons occupy the seven levels
$1g_{9/2}$, $0i_{11/2}$, $0j_{15/2}$, $2d_{5/2}$, $3s_{1/2}$, $1g_{7/2}$, and  $2d_{3/2}$ of the 126-184 shell. 

As regards the choice of the single-proton and single-neutron energies, we have proceeded as follows. For Sb isotopes, we have taken them from the experimental spectra of  $^{133}$Sb and $^{133}$Sn, with the exception of the proton
$s_{1/2}$ and the neutron $i_{13/2}$ levels, which are still missing. The values of 
$\epsilon_{s_{1/2}}$ and $\epsilon_{i_{13/2}}$  have been taken from Refs. \cite{Andreozzi97} and \cite{Coraggio02}, respectively, where it is discussed how they are determined.
In the same way, for Bi isotopes, we have made use of the experimental spectra of 
$^{209}$Bi and $^{209}$Pb. The adopted values of the single-particle energies  are reported in Refs. \cite{Coraggio05} and \cite{Coraggio07} for Sb and Bi isotopes, respectively. 

As already mentioned in the Introduction, in our shell-model calculations we have made use of effective interactions derived from the CD-Bonn $NN$ potential renormalized through the  $V_{\rm low-k}$ procedure. As in our previous studies \cite{Coraggio05,Coraggio06,Covello07,Simpson07,Coraggio07}, the cutoff momentum $\Lambda$ is given  the value 2.2 fm$^{-1}$. The computation  of the diagrams included in the $\hat{Q}$ box is performed within the harmonic-oscillator basis using  intermediate states composed of all possible hole states and particle states restricted to the five proton and neutron shells above the Fermi surface. The oscillator parameter is 7.88 Mev for the $A=132$ region and 6.88 MeV for the $A=208$ region, as obtained from the expression $\hbar \omega=45 A^{-1/3} -25 A^{-2/3}$.

Let us now come to the results of our calculations, which have been performed by using the NushellX shell-model code \cite{Nushell}.

In Fig. 1 we show the energies of the first eight calculated states, which are the members of the $\pi g_{7/2} \nu f_{7/2}$  multiplet, and compare them with the eight lowest-lying experimental states \cite{NNDCU}.  
The wave functions of these states are characterized  by 
very little configuration mixing, the percentage of the leading component having a minimum value of 88\% for the $J^{\pi}=2^{-}$ state 
while ranging from 94\% to 100\% for all other states.

\begin{figure}[h]
\includegraphics[height=.25\textheight]{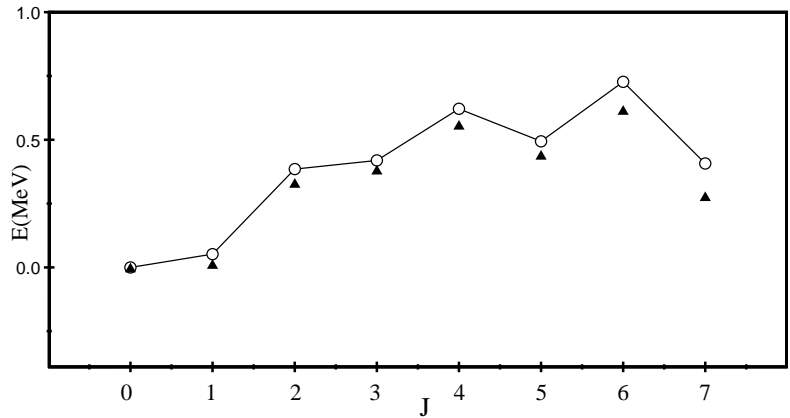}
\caption{Proton-neutron $\pi g_{7/2} \nu f_{7/2}$  multiplet in $^{134}$Sb. The theoretical results are represented by open circles and the experimental data by solid triangles.} 
\end{figure}

As regards $^{136}$Sb, with two more valence neutrons, its ground state was identified as $1^-$ in the early  $\beta$-decay study of Ref. \cite{Hoff97} while the spectroscopic study of Ref. \cite{Mineva01} led to the observation of a $\mu$s isomeric state, which was tentatively assigned a spin and parity of $6^-$. Very recently \cite{Simpson07}, new experimental information has been obtained on the $\mu$s isomeric cascade, leading to the identification of two more excited states.

\begin{figure}
\includegraphics[height=.25\textheight]{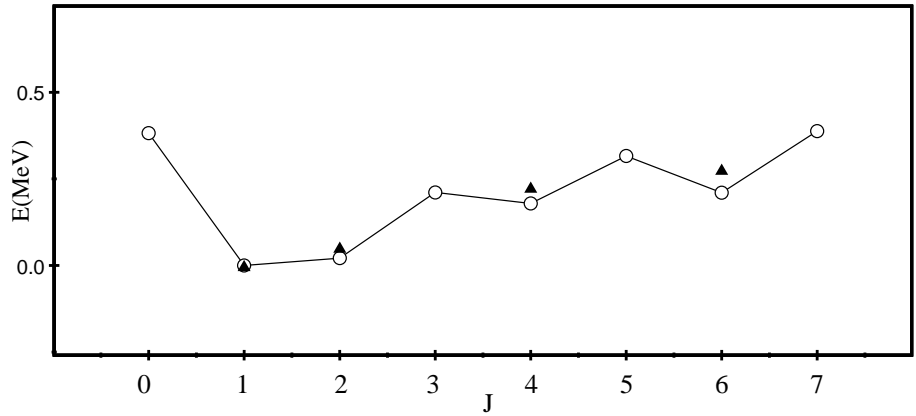}
\caption {Proton-neutron states in $^{136}$Sb arising from the configuration $\pi g_{7/2} \nu (f_{7/2})^{3}$. The theoretical results are represented by open circles and the experimental data by solid triangles.} 
\end{figure}

This achievement was at the origin of our realistic shell-model calculation for this nucleus \cite{Simpson07}, whose results we are now going to present. In Fig.~2, we show the four observed levels together with the calculated yrast states having angular momentum from $0^-$ to $7^-$, which all arise from the $\pi g_{7/2} \nu (f_{7/2})^{3}$ configuration. These states may be viewed as the evolution of the $\pi g_{7/2} \nu f_{7/2}$ multiplet in $^{134}$Sb.

From Figs. 1 and 2 we see that the agreement between theory and experiment is very good, the  discrepancies being in the order of a few tens of keV for most of the states. The largest discrepancy occurs for the $7^{-}$ state in $^{134}$Sb, which lies at about 130 keV 
above its experimental counterpart. It is an important outcome of our calculation for 
$^{134}$Sb that we predict  almost the right spacing between the $0^{-}$ ground state and first excited $1^{-}$ state. In fact, the latter has been observed at 13 keV excitation energy, our value being 53  keV. 

\begin{figure}[h]
\includegraphics[height=0.25\textheight]{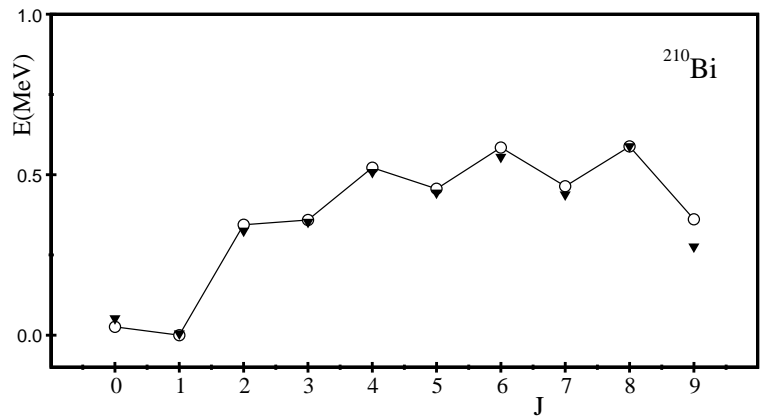}
\caption{Proton-neutron $\pi h_{9/2} \nu g_{9/2}$ 
multiplet in $^{210}$Bi. The theoretical results are represented by open 
circles while the experimental data by solid triangles.}
\end{figure}

In Fig. 3 the calculated $\pi h_{9/2} \nu g_{9/2}$  multiplet in $^{210}$Bi is reported and compared with the experimental data \cite{NNDCE}. This multiplet is the counterpart of the 
$\pi g_{7/2} \nu f_{7/2}$  multiplet in $^{134}$Sb, with the substitution 
$n, l, j \rightarrow n,\, l+1,\, j+ 1$. Comparison of Figs. 1 and 3 shows evidence of the similarity between the behavior of the two multiplets.

Let us now come to $^{212}$Bi, which is the counterpart of $^{136}$Sb. In Fig. 4 we show  the calculated yrast states having angular momentum from $0^-$ to $9^-$, which all arise from the $\pi h_{9/2} \nu (g_{9/2})^{3}$ configuration, and compare them with the available experimental data \cite{NNDCE}. From Fig. 4 we see that the pattern of the multiplet is substantially different from that in $^{210}$Bi, but quite similar to that predicted for  $^{136}$Sb.
In particular, as compared to the case of the one proton-one neutron nuclei $^{134}$Sb and $^{210}$Bi, we see  that the curves of Figs. 3 and 4 are no longer  concave downwards and that  $0^-$ state is raised at about 380 keV and 200 keV excitation energy in $^{136}$Sb and $^{212}$Bi, respectively.

\section{Summary}

We have presented here the results of a shell-model study of neutron-rich nuclei around 
$^{132}$Sn, focusing attention on proton-neutron multiplets in the odd-odd isotopes
$^{134}$Sb and $^{136}$Sb.  We have compared the results obtained for these two nuclei far from stability with those for $^{210}$Bi and $^{212}$Bi, which are their counterparts in the region of stable doubly magic $^{208}$Pb. In both cases, the two-body effective interaction has been derived by means of a $\hat Q$-box folded-diagrams method from the CD-Bonn $NN$ potential, the short-range repulsion of the latter being renormalized by use of the low-momentum potential $V_{\rm low-k}$.

Our results for all four nuclei are in very good agreement with the available experimental data and show evidence of a striking resemblance between the behavior of the multiplets in the $^{132}$Sn and $^{208}$Pb regions. The observation of the missing states of the multiplets in $^{136}$Sb and $^{212}$Bi is certainly needed to further verify the soundness of our findings.

\begin{figure}
\includegraphics[height=0.25\textheight]{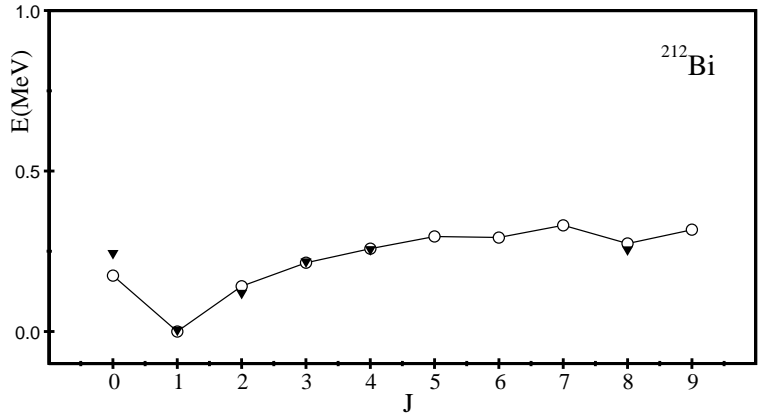}
\caption{Proton-neutron states in $^{212}$Bi arising from the configuration
$\pi h_{9/2} \nu (g_{9/2})^3$. The theoretical results are represented by open 
circles while the experimental data by solid triangles.}
\end{figure}

\begin{theacknowledgments}
This work was supported in part by the Italian Ministero dell'Istruzione, dell'Università e della Ricerca (MIUR).  
\end{theacknowledgments}


\end{document}